\documentstyle[12pt]{article}

\begin{document}

\begin{flushright}
DPUR/TH/70\\
March, 2021\\
\end{flushright}
\vspace{20pt}

\pagestyle{empty}
\baselineskip15pt

\begin{center}
{\large\bf  Quadratic Gravity and Restricted Weyl Symmetry
\vskip 1mm }

\vspace{20mm}

Arata Kamimura\footnote{
           E-mail address:\ k208328@eve.u-ryukyu.ac.jp
                  }
and Ichiro Oda\footnote{
           E-mail address:\ ioda@sci.u-ryukyu.ac.jp
                  }

\vspace{10mm}
           Department of Physics, Faculty of Science, University of the 
           Ryukyus,\\
           Nishihara, Okinawa 903-0213, Japan\\

\end{center}


\vspace{10mm}
\begin{abstract}

We investigate the relationship between quadratic gravity and a restricted Weyl symmetry where a gauge 
parameter $\Omega(x)$ of Weyl transformation satisfies a constraint $\Box \Omega = 0$ in a curved space-time. 
First, we briefly review a model with a restricted gauge symmetry on the basis of QED where a $U(1)$ gauge 
parameter $\theta(x)$ obeys a similar constraint $\Box \theta = 0$ in a flat Minkowski space-time, and 
explain that the restricted gauge symmetry removes one on-shell mode of gauge field, which together with
the Feynman gauge leaves only two transverse polarizations as physical states. Next, it is shown that the restricted 
Weyl symmetry also eliminates one component of a dipole field in quadratic gravity around a flat Minkowski background, 
leaving only a single scalar state.  Finally, we show that the restricted Weyl symmetry cannot remove any dynamical
degrees of freedom in static background metrics by using the zero-energy theorem of quadratic gravity. 
This fact also holds for the Euclidean background metrics without imposing the static condition.  

\end{abstract}

\newpage
\pagestyle{plain}
\pagenumbering{arabic}


\section{Introduction}

We have recently watched a renewed interest in scale invariant gravitational theories since scale symmetries are
expected to play a dominant role at high energies \cite{Shaposhnikov}-\cite{Oda-Higgs}. For instance, let us consider the era of 
Big Bang which is in very high energies. At the Big Bang, it was extremely hot, so the kinetic energy of particles 
would be so enormous as to have completely  overwhelmed the particles' rest mass. In such a situation, 
the rest mass would be effectively irrelevant, that is, as good as zero, so that the universe at early times, could 
be described in terms of effectively massless particles, for which scale symmetries naturally emerge. 
If scale symmetries are exact ones which are only spontaneously broken as in the gauge symmetries 
in the standard model (SM) of particle physics, they might shed some light on various important unsolved problems such as 
cosmological constant problem, the origin of particles' mass and the gauge hierarchy problem etc. 

In this article, we further explore an idea that there might be an intermediate scale symmetry between local (or Weyl) and 
global scale symmetries, which is dubbed a restricted Weyl symmetry in a curved space-time \cite{Edery1}-\cite{Oda-R}. 
In the restricted Weyl symmetry, a gauge transformation parameter, which we will call $\it{scale \, factor}$ $\Omega(x)$, 
is constrained by a harmonic condition $\Box \Omega = 0$ whereas in a conventional or full Weyl transformation 
the scale factor is an unconstrained and free function.   
In the restricted Weyl symmetry, we are allowed to work with actions with dimensionless coupling constants as in globally scale invariant
gravitational theories, which should be contrasted to the situation in the full Weyl symmetry where only the 
conformal (or Weyl) tensor squared is an invariant action if we neglect the other fields except for the metric $g_{\mu\nu}$.

Another advantage of exploring theories with restricted symmetries is that we can understand spontaneous symmetry 
breakdown of global symmetries \cite{Oda-R}. For instance, as explained in the next section, a QED with a restricted gauge symmetry
has been used to understand why the photon is exactly massless by studying the zero-mode solutions to the harmonic
condition on the gauge parameter \cite{Ferrari}. In fact, it has been already shown that a gravitational theory with the restricted Weyl
symmetry naturally gives rise to spontaneous symmetry breakdown of a global scale symmetry \cite{Oda-R}.  

The structure of this article is the following: In Section 2, we review a restricted gauge symmetry in QED which appears 
after we fix the gauge invariance by the Lorenz gauge.  The restricted gauge symmetry resembles the restricted Weyl symmetry 
in the sense that both the symmetries are constrained by an equation $\Box \Theta = 0$ where $\Box$ denotes 
the d'Alembertian operator and $\Theta$ stands for a generic gauge transformation parameter although there is a big difference 
in that in the former case the operator $\Box$ is the d'Alembertian operator in a flat Minkowski space-time whereas 
in the latter case it is defined in a curved Riemannian space-time.\footnote{To distinguish the difference
of the d'Alembertian operators, we will henceforth use $\Box_\eta$ for the Minkowski space-time and $\Box_g$ for the
Riemannian space-time, respectively.} 
In Section 3, we introduce quadratic gravity with the restricted Weyl symmetry and discuss a relation between 
the quadratic gravity and conformal gravity on the basis of the BRST formalism. In Section 4, using conformal gravity,
we explain how the restricted Weyl symmetry can be understood in the framework of the BRST quartet mechanism.
In Section 5, we present a proof that the restricted Weyl symmetry in fact kills one dynamical degree of freedom
in the quadratic gravity. 
In Section 6, the physical meaning of the constraint in the restricted Weyl symmetry is investigated, for which the zero-energy 
theorem of the quadratic gravity \cite{Boulware} gives us information on the falloff behavior at infinity. The final section is 
devoted to conclusion.

\section{Review of restricted gauge symmetry}

In order to clarify the problems which we attempt to understand in this article, it is worthwhile to start with QED 
which is gauge-fixed by the Lorenz gauge whose Lagrangian takes the form
\begin{eqnarray}
{\cal{L}} = - \frac{1}{4} F_{\mu\nu}^2 + \bar \psi ( i \gamma^\mu \partial_\mu - m ) \psi
+ e A_\mu \bar \psi \gamma^\mu \psi + B \partial_\mu A^\mu + \frac{\alpha}{2} B^2,
\label{QED}  
\end{eqnarray}
where $A_\mu, \psi$ and $B$ are respectively the electromagnetic field, spinor field and Nakanishi-Lautrup field,
the field strength is defined as $F_{\mu\nu} \equiv \partial_\mu A_\nu - \partial_\nu A_\mu$, and $\alpha$ is a real number
called ``gauge parameter'' which can be chosen at our disposal. Even if the $U(1)$ gauge symmetry has been already
gauge-fixed by the Lorenz gauge, this Lagrangian still has a residual symmetry called a restricted $U(1)$ gauge 
invariance\footnote{This residual gauge transformation was also called the ``c-number'' gauge transformation long ago 
in comparison to the ``q-number'' gauge transformation where the gauge parameter is also changed \cite{Nakanishi, Yokoyama}.} 
which is explicitly given by
\begin{eqnarray}
\delta A_\mu = \partial_\mu \theta, \quad 
\delta \psi = i e \theta \psi, \quad
\delta \bar \psi = - i e \theta \bar \psi, \quad 
\delta B = 0,
\label{Res-Gauge}  
\end{eqnarray}
where $\Box_\eta \theta = 0$ where $\Box_\eta = \eta^{\mu\nu} \partial_\mu \partial_\nu$ is the d'Alembertian operator
in a flat Minkowski space-time with the flat metric $\eta_{\mu\nu} = \textrm{diag} (-1, 1, 1, 1)$.

The restricted $U(1)$ gauge symmetry has two physically interesting aspects. The one is that using this symmetry
one can understand why the photon is exactly massless. Actually, we can construct the generator corresponding
to the restricted $U(1)$ gauge transformation (\ref{Res-Gauge})
\begin{eqnarray}
Q = \int d^3 x ( \theta \partial_0 B - \partial_0 \theta \cdot B ).
\label{RWS-charge}  
\end{eqnarray}
Based on canonical quantization procedure, we can calculate a four-dimensional commutation relation
\begin{eqnarray}
[ A_\mu (x), B(y) ] = i \partial_\mu D (x-y),
\label{4D-CR}  
\end{eqnarray}
where $D (x)$ is called ``invariant D function'' which is defined as
\begin{eqnarray}
D (x) \equiv - \frac{i}{(2 \pi)^3} \int d^4 k \, \varepsilon (k^0) \delta (k^2) e^{ik \cdot x},
\label{inv-D}  
\end{eqnarray}
and obeys the relations
\begin{eqnarray}
\Box_\eta D(x) = 0, \qquad D (0, \vec{x}) = 0, \qquad \partial_0 D (0, \vec{x}) = - \delta (\vec{x}).
\label{inv-D2}  
\end{eqnarray}
Using Eqs. (\ref{RWS-charge}), (\ref{4D-CR}) and (\ref{inv-D2}), we can derive the following vacuum expectation value:
\begin{eqnarray}
\langle 0 | [ i Q, A_\mu(x) ] | 0 \rangle = \partial_\mu \theta (x),
\label{RWS-VEV}  
\end{eqnarray}
which implies that $Q$ is necessarily broken spontaneously if $\theta (x)$ is not a constant. Then, the Goldstone
theorem tells us that $A_\mu$ contains a massless state, which can be identified with the photon after a detailed analysis
\cite{Ferrari}.  In other words, the restricted gauge symmetry provides us with useful information on spontaneous
symmetry breakdown of global symmetries.\footnote{In the detailed analysis, we need to consider zero-mode solutions
to the constraint $\Box_\eta \theta = 0$, which are explicitly given by $\theta(x) = a_\mu x^\mu + b$ where $a_\mu$ and
$b$ are constants. It is global symmetries corresponding to global parameters $a_\mu$ and $b$ that spontaneous symmetry
breakdown could occur. Precisely speaking, the Ferrari-Piccaso's proof is not completely correct, but the claim is true 
that the required massless particle must be a vector when the U(1) symmetry corresponding to the $\theta(x) = b$ 
transformation remains unbroken.}

The other interesting aspect is that we can gauge away one on-shell mode of $A_\mu$ by means of the symmetry (\ref{Res-Gauge}).
With regards to this point, let us note that since $\Box_\eta \theta = 0$ is nothing but the Klein-Gordon equation for a massless 
real scalar field, a general solution is given by
\begin{eqnarray}
\theta (x) = \int \frac{d^3 k}{\sqrt{(2 \pi)^3 2 k_0}} [ a(k) e^{ikx}
+ a^\dagger(k) e^{-ikx} ],
\label{Omega}  
\end{eqnarray}
where $k_0 = |\vec{k}|$.

The field equations from the Lagrangian (\ref{QED}) read
\begin{eqnarray}
&{}& \partial_\nu F^{\nu\mu} - \partial^\mu B + e j^\mu = 0, 
\nonumber\\ 
&{}& \partial^\mu A_\mu + \alpha B = 0,
\nonumber\\
&{}& [ i \gamma^\mu ( \partial_\mu - i e A_\mu ) - m ] \psi = 0,
\label{QED-Eq}  
\end{eqnarray}
where the electric current $j^\mu$ is defined by $j^\mu = \bar \psi \gamma^\mu \psi$. Using these field equations,
we can derive the following equation
\begin{eqnarray}
\Box_\eta A^\mu - ( 1 - \alpha ) \partial^\mu B + e j^\mu = 0.
\label{Gauge-Eq}  
\end{eqnarray} 
In the absence of the electric current $j^\mu = 0$ and with the Feynman gauge $\alpha = 1$, the gauge field obeys 
the same field equation $\Box_\eta A_\mu = 0$ as in $\Box_\eta \theta = 0$. This fact implies that in terms of 
the restricted gauge symmetry we can gauge away one on-shell mode of the gauge field, for instance, a temporal component 
of the gauge field, $A_0 = 0$, which together with the Feynman gauge condition leaves two physical massless states 
for $A_\mu$ identified with two transverse polarizations of the photon.\footnote{In the Lorenz gauge, this statement obviously
holds, but in the Feynman gauge it is a bit sutble. Namely, in the Feynman gauge, the NL field is determined as 
$B = - \partial^\mu A_\mu$ and it appears in the physical Hilbert space, but it is not observed owing to a zero-norm state, 
leading to the desired result.} 

Our purposes in this article is two-fold, one of which is to investigate whether a similar mechanism for eliminating dynamical 
degrees of freedom works even for the restricted Weyl symmetry or not. We will see that the restricted Weyl symmetry can 
indeed gauge away one component of the metric tensor around a flat Minkowski background as in the restricted gauge symmetry.
The other purpose is to understand the physical meaning of the constraint of the restricted Weyl symmetry. We find that 
in case of static background metrics with the timelike Killing vector, the restricted Weyl symmetry reduces to a global scale 
symmetry by using the falloff condition at infinity which is required by the zero-energy theorem of quadratic gravity \cite{Boulware}. 
It also turns out that this holds for any Euclidean metrics without the static condition.

\section{Quadratic gravity with restricted Weyl symmetry}

In this section, we consider the most general gravitational theory without matter fields which is invariant under 
a restricted Weyl transformation (as well as a global scale transformation) and explain its relation to conformal gravity. 
This gravitational theory, which we will call ${\it{quadratic \, gravity}}$, is described by the Lagrangian\footnote{We follow 
the conventions and notation of the MTW textbook \cite{MTW}.}:
\begin{eqnarray}
{\cal{L}} = \sqrt{-g} \, ( \alpha \, R^2 + \beta \, C_{\mu\nu\rho\sigma}^2 ),
\label{QG-Lag}  
\end{eqnarray}
where $\alpha, \beta$ are dimensionless constants and $C_{\mu\nu\rho\sigma}$ the conformal tensor (or Weyl tensor),
and $R$ the scalar curvature. Note that this Lagrangian is invariant under both a global scale transformation and a restricted 
Weyl transformation, but is not so under a local scale (or Weyl) transformation because of the presence of an $R^2$ term.
The Gauss-Bonnet topological invariant enables us to rewrite (\ref{QG-Lag}) into the form
\begin{eqnarray}
{\cal{L}} = \sqrt{-g} \, ( \alpha \, R^2 + \gamma \, R_{\mu\nu}^2 ),
\label{QG-Lag-GB}  
\end{eqnarray}
where $\gamma$ is also a dimensionless constant. In this case, both $R^2$ and $R_{\mu\nu}^2$ terms break the local scale
symmetry even if they are invariant under the restricted Weyl and global scale transformations.
    
Here the restricted Weyl transformation is defined by the Weyl transformation 
\begin{eqnarray}
g_{\mu\nu} \rightarrow g^\prime_{\mu\nu} = \Omega^2 (x) g_{\mu\nu}, 
\label{Full-Weyl}  
\end{eqnarray}
where the gauge transformation parameter, i.e., scale factor $\Omega(x)$, obeys a harmonic condition $\Box_g \Omega = 0$ 
in a curved space-time \cite{Edery1}-\cite{Oda-R}. Note that in proving the invariance of the Lagrangian (\ref{QG-Lag}) 
under the restricted Weyl transformation, we need to use the transformation law of the scalar curvature under (\ref{Full-Weyl}):
\begin{eqnarray}
R \rightarrow R^\prime = \Omega^{-2} ( R - 6 \Omega^{-1} \Box_g \Omega ). 
\label{R-Weyl}  
\end{eqnarray}

The field equations coming from (\ref{QG-Lag}) read
\begin{eqnarray}
&{}&\alpha \Biggl( R_{\mu\nu} R - \frac{1}{4} g_{\mu\nu} R^2 - \nabla_\mu \nabla_\nu R + g_{\mu\nu} \Box_g R \Biggr)
- \beta \Biggl( \frac{2}{3} R_{\mu\nu} R - \frac{1}{6} g_{\mu\nu} R^2 
\nonumber\\
&{}& + \frac{1}{3} \nabla_\mu \nabla_\nu R + \frac{1}{6} g_{\mu\nu} \Box_g R 
- \Box_g R_{\mu\nu}  - 2 R_\mu \,^\alpha R_{\alpha\nu} + \frac{1}{2} g_{\mu\nu} R_{\alpha\beta}^2 \Biggr) = 0.
\label{Field-eq}  
\end{eqnarray}
Taking the trace of Eq. (\ref{Field-eq}) produces a harmonic equation for the scalar curvature:
\begin{eqnarray}
\Box_g R = 0.
\label{R-eq}  
\end{eqnarray}

Now we wish to clarify a relation between the quadratic gravity (\ref{QG-Lag}) and conformal gravity:
\begin{eqnarray}
{\cal{L}}_c = \sqrt{-g} \, \beta \, C_{\mu\nu\rho\sigma}^2,
\label{CG-Lag}  
\end{eqnarray}
which is the unique locally scale invariant gravitational theory in the absence of matter fields.
In fact, we will see that the Lagrangian (\ref{QG-Lag}) can be obtained from conformal gravity by fixing the local scale symmetry 
in terms of a gauge condition $R = 0$ up to the FP-ghost term \cite{Oda-R}. 

In order to understand this fact, it is useful to make use of the BRST formalism. The BRST transformation for the local scale 
transformation reads
\begin{eqnarray}
\delta_B g_{\mu\nu} &=& 2 c g_{\mu\nu}, \qquad
\delta_B \sqrt{-g} = 4 c \sqrt{-g}, \qquad
\delta_B R = - 2 c R - 6 \Box_g c,
\nonumber\\
\delta_B \bar c &=& i B, \qquad \delta_B c = \delta_B B = 0.
\label{BRST}  
\end{eqnarray}

Then, the Lagrangian for the gauge condition and the FP ghosts can be obtained by a standard method \cite{Kugo-Uehara}:
\begin{eqnarray}
{\cal{L}}_{GF+FP} &=& - i \delta_B \left[ \sqrt{-g} \bar c \left( R + \frac{\alpha}{2} B \right) \right]
\nonumber\\
&=& \sqrt{-g} \left( \hat B R + \frac{\alpha}{2} \hat B^2 - 6 i \bar c \Box_g c \right)
\nonumber\\
&=& \sqrt{-g} \left( - \frac{1}{2 \alpha} R^2 + 6 i g^{\mu\nu} \partial_\mu \bar c \partial_\nu c \right),
\label{GF-FP}  
\end{eqnarray}
where we have defined $\hat B \equiv B + 2 i \bar c c$ and in the last step we performed the
path integral over the auxiliary field $\hat B$ and integration by parts \cite{Oda-R}.

Finally, adding (\ref{GF-FP}) to the classical Lagrangian (\ref{CG-Lag}), we arrive at a gauge-fixed and BRST-invariant 
Lagrangian:
\begin{eqnarray}
{\cal{L}}_q &=& \sqrt{-g} \Biggl( \beta \, C_{\mu\nu\rho\sigma}^2 - \frac{1}{2 \alpha} R^2 + 6 i g^{\mu\nu} 
\partial_\mu \bar c \partial_\nu c \Biggr).
\label{BRST-Lag}  
\end{eqnarray}
Dropping the last FP-ghost term and rewriting $- \frac{1}{2 \alpha}$ as $\alpha$ leads to the Lagrangian
(\ref{QG-Lag}), which is invariant under the restricted Weyl transformation (as well as a global scale transformation).
In this way, we have succeeded in deriving the quadratic gravity (\ref{QG-Lag}) with the restricted Weyl symmetry and the global
scale symmetry by beginning with the conformal gravity (\ref{CG-Lag}) with the local scale symmetry by taking the gauge condition 
$R = 0$.
 
Incidentally, it is of interest to point out that the FP-ghost Lagrangian is also invariant under not the full Weyl 
transformation but the restricted Weyl transformation. For this aim, let us first assume that both FP-ghost 
and FP-antighost have the Weyl weight $-1$, that is, under the Weyl transformation they transform as
\begin{eqnarray}
c \rightarrow c^\prime = \Omega^{-1} (x) c, \qquad 
\bar c \rightarrow \bar c^\prime = \Omega^{-1} (x) \bar c. 
\label{Weyl-Ghost}  
\end{eqnarray}
Then, we find that under the Weyl transformation the ghost term transforms as
\begin{eqnarray}
&{}& \sqrt{-g} g^{\mu\nu} \partial_\mu \bar c \partial_\nu c 
\nonumber\\
&\rightarrow& \sqrt{-g} g^{\mu\nu} \left[ \partial_\mu \bar c \partial_\nu c
+ \Omega^{-1} \nabla_\mu \nabla_\nu \Omega \cdot \bar c c
- \nabla_\mu (\Omega^{-1} \nabla_\nu \Omega \cdot \bar c c) \right]. 
\label{Ghost-transf}  
\end{eqnarray}
Hence, the ghost kinetic term is invariant under the restricted Weyl transformation up to a surface
term.

\section{The BRST quartet mechanism}

The derivation of the BRST-invariant Lagrangian (\ref{BRST-Lag}) from conformal gravity 
(\ref{CG-Lag}) via the gauge fixing procedure demonstrates that the physical content between the two theories is 
exactly the same since Eq. (\ref{GF-FP}) is BRST-exact. In order to explain how the restricted Weyl symmetry can 
be understood in the framework of the BRST formalism, it is useful to check the quantum equivalence between (\ref{CG-Lag}) 
and (\ref{BRST-Lag}) through the BRST quartet mechanism \cite{Kugo-Ojima} in an explicit manner. 

To this aim, let us expand the metric around the flat Minkowski background metric
$\eta_{\mu\nu}$:
\begin{eqnarray}
g_{\mu\nu} = \eta_{\mu\nu} + h_{\mu\nu}, 
\label{Flat-exp}  
\end{eqnarray}
where $h_{\mu\nu}$ indicates a small fluctuation. Moreover, we use the York decomposition for the metric fluctuation
field $h_{\mu\nu}$ \cite{York}:
\begin{eqnarray}
h_{\mu\nu} = h^{TT}_{\mu\nu} + \partial_\mu \xi_\nu +  \partial_\nu \xi_\mu +  \partial_\mu \partial_\nu \sigma
- \frac{1}{4} \eta_{\mu\nu} \Box_\eta \sigma + \frac{1}{4} \eta_{\mu\nu} h, 
\label{York}  
\end{eqnarray}
where $h^{TT}_{\mu\nu}$ is both transverse and traceless, and $\xi_\mu$ is transverse:
\begin{eqnarray}
\partial^\mu h^{TT}_{\mu\nu} = \eta^{\mu\nu} h^{TT}_{\mu\nu} = \partial^\mu \xi_\mu = 0.
\label{York-conditions}  
\end{eqnarray}
From the BRST transformation (\ref{BRST}), we find the following BRST transformation for asymptotic
fields:
\begin{eqnarray}
\delta_B s = 8 c, \qquad
\delta_B c = 0, \qquad
\delta_B \bar c = i \hat B, \qquad 
\delta_B \hat B = 0,
\label{Asymp-BRST}  
\end{eqnarray}
where $s \equiv h - \Box_\eta \sigma$ is invariant under the general coordinate transformation as shown in the next 
section. 

Next, let us consider the field equations for the above asymptotic fields. The scalar curvature can be written
in terms of $s(x)$ at the linear order as
\begin{eqnarray}
R = - \frac{3}{4} \Box_\eta s.
\label{R-s}  
\end{eqnarray}
The field equation (\ref{R-eq}) then produces the field equations for $s(x)$\footnote{Here we can neglect contributions 
from the FP-ghost term since they start from the second order in fields.}  
\begin{eqnarray}
\Box_\eta^2 s = 0,
\label{R-s2}  
\end{eqnarray}
which means that $s(x)$ is a dipole field.  From Eq. (\ref{GF-FP}), we have the field equation for the NL field $\hat B$:
\begin{eqnarray}
\hat B = - \frac{1}{\alpha} R.
\label{R-B}  
\end{eqnarray}
With the help of Eq. (\ref{R-eq}), Eq. (\ref{R-B}) yields
\begin{eqnarray}
\Box_\eta \hat B = 0,
\label{hat-B-eq}  
\end{eqnarray}
which implies that the NL field $\hat B$ is a simple pole field. As for the field equations for the FP-ghosts, from Eq. (\ref{GF-FP})
we can easily find that 
\begin{eqnarray}
\Box_\eta c = \Box_\eta \bar c = 0.
\label{hat-B-eq2}  
\end{eqnarray}
Thus, the FP-ghosts are also simple pole fields. 

The dipole field $s(x)$ can be decomposed into two massless scalar fields. Actually, the dipole equation is 
expressed in terms of two coupled equations by introducing an additional scalar field $\phi(x)$ as
\begin{eqnarray}
\Box_\eta s = \phi, \qquad \Box_\eta \phi = 0.
\label{Coupled-dipole-eq}  
\end{eqnarray}
Then, by introducing a nonlocal operator ${\cal{D}}$ \cite{Kugo-Ojima}, which works as an inverse of the d'Alembertian 
operator $\Box_\eta$, defined by 
\begin{eqnarray}
{\cal{D}} \equiv - \frac{1}{2 \Delta} \left( x_0 \partial_0 - \frac{1}{2} \right),
\label{D-op}  
\end{eqnarray}
we can extract a simple pole field $\tilde s$ as 
\begin{eqnarray}
\tilde s = s - {\cal{D}} \phi.
\label{tilde-s}  
\end{eqnarray}

Using $\tilde s$ instead of $s$, the BRST transformation (\ref{Asymp-BRST}) becomes
\begin{eqnarray}
\delta_B \tilde s = 8 c, \qquad
\delta_B c = 0, \qquad
\delta_B \bar c = i \hat B, \qquad 
\delta_B \hat B = 0,
\label{Asymp-BRST2}  
\end{eqnarray}
and all the asymptotic fields satisfy the simple pole field equation: 
\begin{eqnarray}
\Box_\eta \tilde s = \Box_\eta c = \Box_\eta \bar c = \Box_\eta \hat B = 0,
\label{Asymp-eq2}  
\end{eqnarray} 
which clearly indicates that the set of the asymptotic fields $\{ \tilde s, c, \bar c, \hat B \}$ belongs to a BRST quartet 
so that they drop from physical Hilbert state by the Kugo-Ojima's subsidiary condition \cite{Kugo-Ojima}.\footnote{The
physical Hilbert space ${\cal{H}}_{phys}$ is defined as the quotient space of ${\cal{V}}_{phys}$ defined by
the subsidiary condition $Q_B {\cal{V}}_{phys} = 0$ ($Q_B$ is the BRST charge corresponding to the Weyl symmetry)
with respect to its zero-norm subspace ${\cal{V}}_0$, i.e., ${\cal{H}}_{phys} = \frac{{\cal{V}}_{phys}}{{\cal{V}}_0}$.
The BRST quartet belongs to the zero-norm subspace  ${\cal{V}}_0$.}  Here it is worth noticing
that using Eqs. (\ref{R-s}), (\ref{R-B}) and (\ref{Coupled-dipole-eq}), the NL field $\hat B$ is proportional to
the scalar field $\phi$:
\begin{eqnarray}
\hat B = - \frac{1}{\alpha} R = \frac{3}{4 \alpha} \Box_\eta s = \frac{3}{4 \alpha} \phi.
\label{B-phi}  
\end{eqnarray}
Since both $\tilde s(x)$ and $\phi(x)$ belong to the BRST quartet, the dipole field $s(x)$ completely drops from the 
physical Hilbert space by the Kugo-Ojima's subsidiary condition as expected.
  
An important lesson obtained by means of the BRST quartet mechanism is that it is essential to decompose 
the dipole field into two scalar fields as in Eq. (\ref{Coupled-dipole-eq}), both of which can be eliminated from 
the physical Hilbert space by the Kugo-Ojima's subsidiary condition \cite{Kugo-Ojima}.  In the quadratic gravity
with the restricted Weyl symmetry discussed in the next section, it turns out that it is also important to make
use of this decomposition even if there is no BRST symmetry.

\section{Restricted Weyl symmetry}

In this section, we will prove that the restricted Weyl symmetry can remove one dynamical degree
of freedom included in the dipole field $s(x)$.

To do so, let us begin by the Lagrangian (\ref{QG-Lag}) and rewrite it in terms of the York variables up to quadratic terms in the fields:
\begin{eqnarray}
{\cal{L}} = \frac{9}{16} \alpha \, s \Box_\eta^2 s + \frac{1}{2} \beta \, h^{TT \mu\nu} \Box_\eta^2 h^{TT}_{\mu\nu}.
\label{QG-Lag-York}  
\end{eqnarray}
The fields equations are therefore given by two dipole equations: 
\begin{eqnarray}
\Box_\eta^2 s = 0, \qquad \Box_\eta^2 h^{TT}_{\mu\nu} = 0.
\label{2-dipole-eq}  
\end{eqnarray}

Under the infinitesimal general coordinate transformation, $h_{\mu\nu}$ transforms as
\begin{eqnarray}
h_{\mu\nu} \rightarrow h_{\mu\nu}^\prime = h_{\mu\nu} + \partial_\mu \varepsilon_\nu
+ \partial_\nu \varepsilon_\mu,
\label{GCT}  
\end{eqnarray}  
where $\varepsilon_\mu$ is the infinitesimal gauge parameter. Here let us divide $\varepsilon_\mu$
into transverse and longitudinal modes:
\begin{eqnarray}
\varepsilon_\mu = \varepsilon_\mu^T + \partial_\mu \varepsilon,
\label{gauge-parameter}  
\end{eqnarray}  
where $\partial^\mu \varepsilon_\mu^T = 0$.
Then, each component in the York decomposition (\ref{York}) transforms like
\begin{eqnarray}
h_{\mu\nu}^{\prime \, TT} &=&  h_{\mu\nu}^{TT}, \qquad 
\xi_\mu^\prime = \xi_\mu + \varepsilon_\mu^T, 
\nonumber\\
\sigma^\prime &=& \sigma + 2 \varepsilon, \qquad
h^\prime = h + 2 \Box_\eta \varepsilon.
\label{Comp-GCT}  
\end{eqnarray}  
Hence, it is obvious that the Lagrangian (\ref{QG-Lag-York}) is manifestly invariant under the general coordinate
transformation owing to $s^\prime = s$.
 
Under the infinitesimal restricted Weyl transformation (\ref{Full-Weyl}), we have
\begin{eqnarray}
h_{\mu\nu}^{\prime \, TT} =  h_{\mu\nu}^{TT}, \qquad 
s^\prime = s + 8 \Lambda, 
\label{Inf-RWT}  
\end{eqnarray}
where we have put the gauge parameter to be its infinitesimal form, $\Omega(x) = e^{\Lambda(x)}$ ($|\Lambda| \ll 1$), 
and $\Lambda$ also obeys a harmonic condition $\Box_\eta \Lambda = 0$. 

Now let us focus our attention to the dipole equation for $s(x)$ in (\ref{2-dipole-eq}). 
As mentioned in Eq. (\ref{Coupled-dipole-eq}), the dipole field $s(x)$ can be expressed in terms of two coupled equations 
by introducing a new scalar field $\phi(x)$. By using the Green's function $G(x, y)$ satisfying $\Box_\eta G(x, y) 
= \delta^4 (x-y)$, a general solution for the former equation in (\ref{2-dipole-eq}) is provided  by\footnote{This equation is 
in essence equivalent to Eq. (\ref{tilde-s}).} 
\begin{eqnarray}
s(x) = \int d^4 y \,  G(x, y) \phi(y) + s_0(x),
\label{s-sol}  
\end{eqnarray}
where $s_0(x)$ satisfies the d'Alembert's equation $\Box_\eta s_0 = 0$. The infinitesimal restricted Weyl transformation 
for the two scalar fields $\phi$ and $s_0$ becomes
\begin{eqnarray}
\phi^\prime = \phi, \qquad 
s^\prime_0 =  s_0 + 8 \Lambda. 
\label{two-RWT}  
\end{eqnarray}
It is then possible to take a gauge condition for the restricted Weyl symmetry 
\begin{eqnarray}
s_0 (x) = 0, 
\label{s0-gauge}  
\end{eqnarray}
since $s_0 (x)$ is a gauge-variant field and both $s_0 (x)$ and $\Lambda(x)$ are harmonic functions.
As a result, the $R^2$ term in the quadratic gravity propagates just a single spin-0 state associated with the 
massless scalar field $\phi(x)$ around a flat Minkowski background, and the field $s(x)$ is described by $\phi(x)$ as
\begin{eqnarray}
s(x) = \int d^4 y \,  G(x, y) \phi(y).
\label{s-sol2}  
\end{eqnarray}
Note that unlike the case of conformal gravity in the previous section, in the quadratic gravity we have no BRST transformation 
for the Weyl transformation, so the massless scalar field $\phi(x)$ in $s(x)$ remains as a propagating state while the scalar
field $s_0(x)$ is eliminated by the restricted Weyl symmetry.

\section{Restricted Weyl symmetry and zero-energy theorem of quadratic gravity}

The d'Alembertian operator in a curved space-time is a geometrical operator in the sense that its definition
needs information on geometry through the metric tensor. It is therefore natural to conjecture that the restricted
Weyl symmetry might not be able to remove dynamical degrees of freedom for certain background geometries. Indeed,
it is well known that there are only static, asymptotically constant solutions to the massless Klein-Gordon equation 
for a scalar field if the first derivative of the scalar field approaches zero sufficiently fast at infinity \cite{Schmidt}-\cite{Kehagias}. 
In this section, we wish to show explicitly that our conjecture is valid if the space-time is static in the sense that 
there is the timelike Killing vector field. It is of interest to note that the falloff conditions necessary for our derivation
stems from a proof of the zero-energy theorem of the quadratic gravity \cite{Boulware}. Moreover, we will show that 
this fact holds for the Euclidean metrics as well although the space-time is not static.

We would like to consider a geometry where the space-time is static in the sense that ${\cal{L}}_t g_{\mu\nu} = 0$
for which $t_\mu$ is the timelike Killing vector field. For clarity, let us consider the following static line element:
\begin{eqnarray}
d s^2 \equiv g_{\mu\nu} d x^\mu d x^\nu = - f^2 (x^k) d t^2 + g_{ij} (x^k) d x^i d x^j,
\label{line}  
\end{eqnarray}
where $f$ and $g_{ij}$ are functions of only spatial coordinates $x^k$ and $i, j, k, \cdots = 1, 2, 3$.

With these assumptions, we multiply the constraint $\Box_g \Omega = 0$ by $f \Omega$ and integrate over
spatial coordinates:
\begin{eqnarray}
\int d^3 x \sqrt{{}^{(3)} g} \, f \Omega \Box_g \Omega = 0,
\label{Identity}  
\end{eqnarray}
where ${}^{(3)} g \equiv \det g_{ij}$. It turns out that $\Box_g \Omega = 0$ can be rewritten into the form\footnote{Here
we have considered that the scale factor is also a function of spatial coordinates, i.e., $\Omega = \Omega(x^k)$ since
a locally scaled metric $g^\prime_{\mu\nu} = \Omega^2(x) g_{\mu\nu}$ must be static as well.}
\begin{eqnarray}
\Box_g \Omega = \frac{1}{f} g^{ij} \partial_i f \partial_j \Omega + g^{ij} D_i D_j \Omega,
\label{Box-Omega}  
\end{eqnarray}
where $D_i$ is the covariant derivative with respect to the spatial metric $g_{ij}$. Substituting Eq. (\ref{Box-Omega})
into Eq. (\ref{Identity}) leads to
\begin{eqnarray}
\int d^3 x \sqrt{{}^{(3)} g} \left[ D^i (f \Omega D_i \Omega) - f  (D_i \Omega)^2 \right] = 0.
\label{Identity2}  
\end{eqnarray}
If $D_i \Omega \rightarrow 0$ sufficiently fast at infinity, then the surface term makes no contribution, and 
the non-positivity of the remaining term produces $D_i \Omega = 0$, which means that the scale factor $\Omega$ is
a constant, thereby the restricted Weyl symmetry reducing to the global scale symmetry.

Next, let us show that $D_i \Omega \rightarrow 0$ sufficiently fast at infinity on the basis of the zero-energy theorem of 
quadratic gravity \cite{Boulware}. In the quadratic gravity, the definition of asymptotic flatness permitting a proof of
the zero-energy theorem is given by \cite{Boulware}
\begin{eqnarray}
g_{ij} \rightarrow \delta_{ij} + {\cal{O}} \left(r^{-1 - \varepsilon} \right),
\label{Asym-falloff}  
\end{eqnarray}
where $\varepsilon$ is a small positive quantity. Since the zero-energy theorem should hold for both a metric $g_{\mu\nu}$
and its locally scaled metric $g^\prime_{\mu\nu} = \Omega^2(x) g_{\mu\nu} = e^{2 \Lambda(x)} g_{\mu\nu}$, the metric
$g^\prime_{\mu\nu}$ must obey the same falloff behavior as $g_{\mu\nu}$. Using an infinitesimal gauge parameter
$\Lambda(x)$, we obtain
\begin{eqnarray}
g^\prime_{ij} &\simeq& ( 1 + 2 \Lambda ) \left[ \delta_{ij} + {\cal{O}} \left(r^{-1 - \varepsilon} \right) \right]
\nonumber\\
&=& \delta_{ij} + 2 \Lambda \delta_{ij} + {\cal{O}} \left(r^{-1 - \varepsilon} \right).
\label{g-falloff}  
\end{eqnarray}
Thus, $\Lambda(x)$ must have at least the asymptotic behavior
\begin{eqnarray}
\Lambda = {\cal{O}} \left(r^{-1 - \varepsilon} \right).
\label{Lambda-falloff}  
\end{eqnarray}
Then, we have
\begin{eqnarray}
D_i \Omega = {\cal{O}} \left(r^{-2 - \varepsilon} \right).
\label{D-Lambda-falloff}  
\end{eqnarray}
Hence, at infinity $D_i \Omega$ goes to zero so rapidly that the surface term in (\ref{Identity2}) vanishes.

Of course, the argument done thus far cannot be applied to the case where there is no timelike Killing
vector. However, in case of the positive semi-definite Euclidean metrics, we can show a similar result without imposing 
the static condition as follows: Start with the constraint $\Box_g \Omega = 0$ of the restricted Weyl transformation, multiply it 
by $\Omega$, and then integrate over space-time coordinates with a four dimensional Euclidean measure $\sqrt{g}$ as
\begin{eqnarray}
0 = \int d^4 x \sqrt{g} \, \Omega \Box_g \Omega = \int d^4 x \sqrt{g}  \left[ \nabla^\mu ( \Omega \nabla_\mu \Omega) 
-  (\nabla_\mu \Omega)^2 \right].
\label{Vol-integral}  
\end{eqnarray}
If $\nabla_\mu \Omega \rightarrow 0$ sufficiently fast at infinity, then the surface term can be dropped, and 
consequently we obtain
\begin{eqnarray}
\nabla_\mu \Omega = 0.
\label{Const-Omega}  
\end{eqnarray}
This equation implies that $\Omega$ is a constant so that the restricted Weyl symmetry reduces to the global scale
symmetry. This fact might be understood from the simplest example, the flat Euclidean metric $\eta_{\mu\nu}
= \delta_{\mu\nu} \equiv \textrm{diag} (+1, +1, +1, +1)$. In this specific case, the momenta $k_\mu$ identically vanish since $k_\mu^2 =0$
simply means $k_\mu = 0$ so that we cannot obtain an infinite number of oscillator solutions as in Eq. (\ref{Omega})
but we have only the constant solution $\Omega = b$ with $b$ being a constant.\footnote{Note that $\Omega = a_\mu x^\mu$
($a_\mu$ is a constant vector) is not a solution since it does not satisfy the condition $\nabla_\mu \Omega \rightarrow 0$ 
at infinity. However, in the flat Euclidean space-time  $g_{\mu\nu} = \delta_{\mu\nu}$, this falloff condition might be too strong 
since $\Omega = a_\mu x^\mu$ corresponds to the special conformal group.}

From the viewpoint of the conventional QFT, it might appear to be strange at first sight that the possibility 
of eliminating dynamical degrees of freedom by using the restricted Weyl symmetry depends on the property
of the background metrics at infinity. In order to account for this issue, let us suppose that the infinitesimal gauge 
parameter $\Lambda$ has the falloff behavior like
\begin{eqnarray}
\Lambda = {\cal{O}} \left(r^{+1 - \varepsilon} \right),
\label{Lambda-falloff2}  
\end{eqnarray}
instead of Eq. (\ref{Lambda-falloff}). Note that with this gauge parameter $\Lambda$, the restricted Weyl transformation
becomes ill-defined at infinity owing to its divergence, so we cannot require the invariance under such a singular
gauge transformation to define the Hilbert space in QFT. To put it differently, using such the singular gauge transformation,
we cannot eliminate any dynamical degrees of freedom existing at infinity. Since gauge symmetries should in general
remove dynamical degrees of freedom over the whole space-time, the singular gauge transfomations are not allowed
as the genuine gauge symmetries.\footnote{Of course, it is possible to consider asymptotic gauge symmetries holding only
at infinity.}

\section{Conclusion}

In this article, we have studied the quadratic gravity, which is invariant under the restricted Weyl transformation as
well as the global scale transformation. It has been explicitly shown that the dipole field $s(x)$, which is involved in the
$R^2$ term, has only a single scalar state from the restricted Weyl symmetry. This mechanism of the elimination of 
a dynamical degree of freedom in the quadratic gravity is very similar to that of the QED with the restricted gauge symmetry. 
Since both the theories also share a similar mechansim of spontaneous symmetry breakdown of global symmetries,
the quadratic gravity might be regarded as a gravitational analog of the QED which is gauged-fixed by the Lorenz gauge. 

Moreover, we have investigated a relation between the quadratic gravity and conformal gravity, which is the unique
gravitational theory with the local scale symmetry, by fixing the Weyl symmetry by the gauge condition $R = 0$. In case of
the absence of the FP-ghost term, the two theories have the same Lagrangian.

We also have investigated the physical meaning of the constraint $\Box_g \Omega = 0$ of the restricted Weyl transformation.
What we have shown is that when the background metrics are static, the restricted Weyl transformation reduces to 
the global scale transformation, thereby making it for the restricted Weyl symmetry impossible to remove a dynamical 
degree of freedom. This observation has been also generalized to the case where the space-time metric has the Euclidean signature.
Of course, the scale factor is in general a function of all the space-time coordinates $x^\mu$ and the space-time metric
does not have the Euclidean signature but specific cases such as the instantons, so the restricted Weyl symmetry 
can indeed eliminate one degree of freedom except for static space-times. Nevertheless, it is of interest to notice 
that the constraint $\Box_g \Omega = 0$ provides us with a different physical meaning depending on the property 
of the space-time.

\begin{flushleft}
{\bf Acknowledgements}
\end{flushleft}

We would like to thank K. Taniguchi and K. Uryu for useful discussions, and in particular, T. Kugo for valuable discussions 
and reading of this manuscript.



\begin{thebibliography}{99}

\bibitem{Shaposhnikov}
M. Shaposhnikov and D. Zenhausern, {``Quantum Scale Invariance, Cosmological Constant and 
Hierarchy Problem", Phys. Lett. {\bf B 671} (2009) 162.}

\bibitem{Oda-H}
I. Oda, {``Higgs Mechanism in Scale-Invariant Gravity'', Adv. Stud. Theor. Phys. {\bf 8} (2014) 
215.}

\bibitem{Alvarez}
L. Alvarez-Gaume, A. Kehagias, C. Kounnas and and D. L\"{u}st, {``Aspects of Quadratic Gravity", Fortschr. Phys. 
{\bf 64} (2016) 176.}

\bibitem{Oda-C}
I. Oda, {``Classical Weyl Transverse Gravity'', Eur. Phys. J. {\bf C 77} (2017) 284.}

\bibitem{Salvio}
A. Salvio, {``Quadratic Gravity", Front. in Phys. {\bf 6} (2018) 77.}

\bibitem{Ghilencea}
D. M. Ghilencea, {``Spontaneous Breaking of Weyl Quadratic Gravity to Einstein Action and Higgs Potential'', 
JHEP {\bf 1903} (2019) 049.}

\bibitem{Oda-P}
I. Oda, {``Planck Scale from Broken Local Conformal Invariance in Weyl Geometry'', PoS CORFU2019 (2020) 070.}

\bibitem{Kugo-N}
T. Kugo, {``Necessity and Insufficiency of Scale Invariance for Solving Cosmological Constant Problem'', 
PoS CORFU2019 (2020) 071.}

\bibitem{Oda-Higgs}
I. Oda, {``Higgs Potential from Weyl Conformal Gravity'', Mod. Phys. Lett. {\bf A 35} (2020) 2050304.}

\bibitem{Edery1}
A. Edery and Y. Nakayama, {``Restricted Weyl Invariance in Four-dimensional Curved Spacetime'',
Phys. Rev. {\bf D 90} (2014) 043007.}

\bibitem{Edery2}
A. Edery and Y. Nakayama, {``Generating Einstein Gravity, Cosmological Constant and Higgs Mass
from Restricted Weyl Invariance'', Mod. Phys. Lett. {\bf A 30} (2015) 1550152.}

\bibitem{Edery3}
A. Edery and Y. Nakayama, {``Critical Gravity from Four Dimensional Scale Invariant Gravity'', 
JHEP {\bf 1911} (2019) 169.}

\bibitem{Oda-R}
I. Oda, {``Restricted Weyl Symmetry'', Phys. Rev. {\bf D 102} (2020) 045008.}

\bibitem{Ferrari}
R. Ferrari and L. E. Picasso, {``Spontaneous Breakdown in Quantum Electrodynamics", 
Nucl. Phys. {\bf B 31} (1971) 316.}  

\bibitem{Boulware}
D. G. Boulware, G. T. Horowitz and A. Strominger, {``Zero Energy Theorem for Scale Invariant Gravity", 
Phys. Rev. Lett. {\bf 50} (1983) 1726.}

\bibitem{Nakanishi}
N. Nakanishi and I. Ojima, {``Covariant Operator Formalism of Gauge Theories and Quantum Gravity'', 
World Scientific, Singapore, 1990.}

\bibitem{Yokoyama}
K. Yokoyama, {``Canonical Quantum Electrodynamics with Invariant One-parameter Gauge Families", 
Prog. Theor. Phys. {\bf 51} (1974) 1956.}   

\bibitem{MTW}
C. W. Misner, K. S. Thorne and J. A. Wheeler, {``Gravitation", W H Freeman and Co (Sd), 1973.}

\bibitem{Kugo-Uehara}
T. Kugo and S. Uehara, {``General Procedure of Gauge Fixing Based on BRS Invariance Principle", 
Nucl. Phys. {\bf B 197} (1982) 378.}  

\bibitem{Kugo-Ojima}
T. Kugo and I. Ojima, {``Local Covariant Operator Formalism of Non-Abelian Gauge Theories and 
Quark Confinement Problem", Prog. Theor. Phys. Suppl. {\bf 66} (1979) 1.}  

\bibitem{York}
J. W. York, {``Conformally Invariant Orthogonal Decomposition of Symmetric Tensors on
Riemannian Manifolds and the Initial Value Problem of General Relativity'', J. Math. Phys. 
{\bf 14} (1973) 456.}

\bibitem{Schmidt}
H.-J. Schmidt, {``On Static Asymptotically Flat Solutions of Fourth Order Gravitational Field Equations'', 
Annalen Phys. {\bf 499} (1987) 361.}

\bibitem{Nelson}
W. Nelson, {``Static Solutions for Fourth Order Gravity'', Phys. Rev. {\bf D 82} (2010) 104026.}

\bibitem{Lu}
H. L\"{u}, A. Perkins, C. N. Pope and K. S. Stelle, {``Black Holes in Higher Derivative Gravity'', 
Phys. Rev. Lett.  {\bf 114} (2015) 171601.}

\bibitem{Kehagias}
A. Kehagias, C. Kounnas, D. D. L\"{u}st and A. Riotto, {``Black Hole Solutions in $R^2$ Gravity'', 
JHEP {\bf 05} (2015) 143.}

 	
 



\end{thebibliography}
\end{document}